%% file: Final_regmodBPTSPonecolumn.tex
\documentclass[11pt,draftcls,onecolumn]{IEEEtran}  
\usepackage{epsfig,graphics,color,subfigure,wrapfig,hyperref,url}
\usepackage{amssymb,amsmath,algorithm,times,caption}
\usepackage{slashbox}
\begin{document}
\input{zcom}

\newcommand{\tx}{\tilde{x}}
\newcommand{\tN}{\tilde{N}}
\newcommand{\stbad}{k_{0}}
\newcommand{\stb}{k_{b}}
\newcommand{\Section}[1]{  \vspace{-0.1in} \section{#1}  \vspace{-0.075in} } 
\newcommand{\Subsection}[1]{ \vspace{-0.1in} \subsection{#1}  \vspace{-0.075in} }
\setlength{\arraycolsep}{1mm}

\newcommand{\muhat}{\hat{\mu}}
\newcommand{\Sp}{\check{s}}



\title{Exact Reconstruction Conditions for Regularized Modified Basis Pursuit\thanks{ Copyright (c) 2012 IEEE. Personal use of this material is permitted. However, permission to use this material for any other purposes must be obtained from the IEEE by sending a request to pubs-permissions@ieee.org.}
\thanks{
A part of this work was presented at Asilomar, 2010 \cite{regmodBP}. This work was partly supported by NSF grant CCF-0917015. The authors are with the Department of Electrical and Computer Engineering, Iowa State University, Ames,
IA 50010 USA (Email: luwei@iastate.edu; namrata@iastate.edu. Phone: 515-294-4012. Fax: 515-294-8432.)
}}
\author{Wei Lu and Namrata Vaswani}
\maketitle

\vspace{-0.5in}

\begin{abstract}
In this work, we obtain sufficient conditions for exact recovery of regularized modified basis pursuit (reg-mod-BP) and discuss when the obtained conditions are weaker than those for modified compressive sensing or for basis pursuit (BP). The discussion is also supported by simulation comparisons. Reg-mod-BP provides a solution to the sparse recovery problem when both an erroneous estimate of the signal's support, denoted by $T$, and an erroneous estimate of the signal values on $T$ are available. 
\end{abstract}

\Section{Introduction}
In this work, we obtain sufficient conditions for exact recovery of regularized modified basis pursuit (reg-mod-BP) and discuss when the obtained conditions are weaker than those for modified compressive sensing \cite{modcsjournal} or for basis pursuit (BP) \cite{BPDN,decodinglp}. Reg-mod-BP was briefly introduced in our earlier work \cite{modcsjournal} as a solution to the sparse recovery problem when both an erroneous estimate of the signal's support, denoted by $T$, and an erroneous estimate of the signal values on $T$, denoted by $(\hat{\mu})_T$, are available. The problem is precisely defined in Sec. I-A.  Reg-mod-BP, given in (\ref{rmc}), tries to find a vector that is sparsest outside the set $T$ among all solutions that are close enough to $(\hat{\mu})_T$ on $T$ and satisfy the data constraint.
In practical applications, $T$ and $(\hat{\mu})_T$ may be available from prior knowledge, or in recursive reconstruction applications, e.g. recursive dynamic MRI \cite{LSCS,modcsjournal}, recursive compressive sensing (CS) based video compression \cite{CSvideosampling,multiscaleCSvideo}, or recursive projected CS (ReProCS) \cite{rrpcp_arxiv,rrpcp_isit} based video layering,
one can use the support and signal estimate from  the previous time instant for this purpose. 


Basis pursuit (BP) was introduced in \cite{BPDN} as a practical (polynomial complexity) solution to the problem of reconstructing a sparse $m \times 1$ vector, $x$, with support denoted by $N$, from an $n \times 1$ measurements' vector, $y \defn Ax$, when $n < m$. BP solves the following convex (actually linear) program:
\begin{equation}
\min_\beta \|\beta\|_1   \text{ subject to } y = A \beta  \label{cs}
\end{equation}
The recent CS literature has provided strong exact recovery results for BP that are either based on the restricted isometry property (RIP) \cite{decodinglp,candes_rip} or that use the geometry of convex polytopes to obtain ``exact recovery thresholds" on the $n$ needed for exact recovery with high probability \cite{donoho,Dohohowl1}. BP is often just referred to as CS in recent works and our work also occasionally does this.

In recent work \cite{modcsjournal}, we introduced the problem of sparse reconstruction with partial and partly erroneous support knowledge, denoted by $T$, and proposed a solution called modified compressive sensing (mod-CS). We obtained exact reconstruction conditions for mod-CS and showed when they are weaker than those for BP. 
Mod-CS tries to find the solution that is sparsest outside the set $T$ among all solutions of $y=A\beta$, i.e. it solves
\begin{equation}
\min_\beta \|\beta_{T^c}\|_1    \text{ subject to }  y=A\beta \label{mc}
\end{equation}
Ideally the above should be referred to as mod-BP, but since we used the term mod-CS when we introduced it, we will retain it here.
Similar problems were also studied in parallel work by von Borries et al.~\cite{camsap07} and Khajehnejad et al.~\cite{weightedl1}. In \cite{weightedl1}, the authors assumed a probabilistic prior on the support, solved the following weighted $\ell_1$ problem and obtained exact recovery thresholds similar to those in \cite{Dohohowl1}:
 \begin{equation}
\min_{\beta}  \|\beta_{T^c}\|_1+\gamma \|\beta_T\|_1    \text{ subject to }  y=A\beta \label{wl1}
\end{equation}
In another related work \cite{iterativesuppdetect}, Wang et al.~showed how to iteratively improve recovery of a {\em single} signal by solving BP in the first iteration, obtaining a support estimate, solving (\ref{mc}) with this support estimate and repeating this. They also obtained exact recovery guarantees for a single iteration.

Another related idea is {\em CS-diff or CS-residual} which recovers the residual signal $x-\hat{\mu}$ by solving (\ref{cs}) with $y$ replaced by $y-A\hat{\mu}$. This is related to our earlier least squares CS-residual (LS-CS) and Kalman filtered CS (KF-CS) ideas \cite{LSCS,kfcsicip}. However, as explained in \cite{modcsjournal}, the residual signals using all these methods have a support size that is equal to or slightly larger than that of $x$ (except if $(\hat{\mu})_T=x_T$). As a result, these do not achieve exact recovery with fewer measurements. The limitations of some other variants of this are also discussed in detail in \cite{regmodBPDN}.  Reg-mod-BP may also be interpreted as a Bayesian or a model-based CS approach. Recent work in this area includes \cite{modelCS,schniter,schniter_hmmtree}.

%

This paper is organized as follows. We introduce reg-mod-BP in Sec. II. In Sec III, we obtain the exact reconstruction result, discuss its implications and give the key lemmas leading to its proof. Simulation comparisons are given in Sec. IV and conclusions in Sec. V.

\newcommand{\sgn}{\text{sgn}}

\Subsection{Notation and Problem Definition}
For a set $T$,  $T^c = \{ i \in [1,...,m], i \notin T \}$. $\emptyset$ is the empty set. We use $|.|$ to denote the cardinality of a set. The same notation is also used for the absolute value of a scalar. The meaning is clear from context.

For a vector $b$, $(b)_T$, or just $b_T$, denotes a sub-vector containing the elements of $b$ with indices in $T$. $\|b\|_k$ means the $\ell_k$ norm of the vector $b$. The notation $b \succeq 0$  ($b \succ 0$) means that each element of the vector $b$ is greater than or equal to (strictly greater than) zero. Similarly $b \preceq 0$ ($b \prec 0$) means each element is less than or equal to (strictly less than) zero. We define the sign pattern, $\text{sgn}(b)$ as:
\begin{eqnarray}
[\text{sgn}(b)]_i= \left \{
\begin{array}{cc}
    b_i/|b_i|\quad & \text{if}\ \  b_i \neq 0 \\
    0 \quad &  \text{if} \ \ b_i=0
  \end{array} \right.
\end{eqnarray}
We use $'$ for matrix transpose. For a matrix $A$, $A_T$ denotes the sub-matrix containing the columns of $A$ with indices in $T$. Also, $\|A\|:= \max_{x \neq 0} \frac{\|Ax\|_2}{\|x\|_2}$ is the induced 2 norm.

Our goal is to solve the sparse reconstruction problem, i.e. reconstruct an $m$-length sparse vector, $x$, with support, $N$, from an $n< m$ length measurement vector,
\bea
y \defn Ax
\eea
when an erroneous estimate of the signal's support, denoted by $T$; and an erroneous estimate of the signal values  on $T$, denoted by $(\hat{\mu})_T$, are available.
The support estimate, $T$, can be rewritten as
\bea
\label{defDeltas}
T \se N \cup \Delta_e \setminus \Delta, \ \text{where}  \
\Delta \defn N \setminus T \ \text{and} \ \Delta_e \defn T \setminus N
\eea
are the errors ($\Delta$ contains the misses while $\Delta_e$ contains the extras) in the support estimate.

The signal value estimate is assumed to be zero along $T^c$, i.e., $$\hat{\mu} = \left[
              \begin{array}{c}
                (\hat{\mu})_T \\
                \mathbf{0}_{T^c} \\
              \end{array}
            \right]$$
and it satisfies
\begin{eqnarray}\label{signalesterror}
(\muhat)_{T} \se (x)_{T} + \nu, \ \text{with} \ \|\nu\|_\infty \le \rho
\label{feasiblecons}
\end{eqnarray}

The {\em restricted isometry constant (RIC) \cite{decodinglp}, $\delta_s$}, for $A$, is defined as the smallest positive real number satisfying $(1- \delta_s) \|c\|_2^2 \le \|A_S c\|_2^2 \le (1 + \delta_s) \|c\|_2^2$ for all subsets $S$ of cardinality $|S| \le s$ and all real vectors $c$ of length $|S|$. The {\em restricted orthogonality constant (ROC) \cite{decodinglp}, $\theta_{s_1,s_2}$}, is defined as the smallest positive real number satisfying $| {c_1}'{A_{T_1}}'A_{T_2} c_2 | \le \theta_{s_1,s_2} \|c_1\|_2 \|c_2\|_2$ for all disjoint sets $T_1, T_2 $ with $|T_1| \le s_1$, $|T_2| \le s_2$ and  $s_1+s_2 \le m$, and for all vectors $c_1$, $c_2$ of length $|T_1|$, $|T_2|$ respectively.
Both {\em $\delta_s$ and $\theta_{s_1,s_2}$ are non-decreasing functions of $s$ and of $s_1$, $s_2$ respectively} \cite{decodinglp}.

We will frequently use the following functions of the RIC and ROC of $A$ in Sec. III:
\bea
\label{def_a}
a_{k}(s,\Sp) \sdefn \frac{\theta_{\Sp,s} + \frac{\theta_{\Sp,k} \ \theta_{s,k}}{1 - \delta_{k}}}{ 1-\delta_s - \frac{\theta_{s,k}^2}{1 - \delta_{k}} } \\
\label{def_K}
K_{\st}(\sd) \sdefn \frac{ \sqrt{1 + \delta_{\sd}} }{1-\delta_{\sd} - \frac{\theta_{\sd,\st}^2}{1 - \delta_{\st}} }
\eea


For the matrix $A$, and for any set $S$ for which ${A_S}'A_S$ is full rank, we define the matrix $M(S)$ as
\bea
M(S) \defn I-A_{S} ({A_{S}}' A_{S})^{-1}{A_{S}}'
\label{defM}
\eea

\Section{Regularized Modified Basis Pursuit}
Mod-CS given in (\ref{mc}) puts no cost on $\beta_T$ and no explicit constraint except $y=A\beta$. Thus, when very few measurements are available, $\beta_T$ can become larger than required in order to satisfy $y=A\beta$ with the smallest $\|\beta_{T^c}\|_1$. A similar, though less, bias will also occur with (\ref{wl1}) when $\gamma < 1$. However, if a signal value estimate on $T$, $(\hat\mu)_T$, is also available, one can use that to constrain $\beta_T$. One way to do this, as suggested in \cite{modcsjournal}, is to add $\lambda \|\beta_T-\hat{\mu}_T\|_{2}^2$ to the mod-CS cost. However, as we saw from simulations, while this does achieve lower reconstruction error, it cannot achieve exact recovery with fewer measurements (smaller $n$) than mod-CS \cite{modcsjournal}. The reason is it puts a cost on the entire $\ell_2$ distance from $(\muhat)_T$ and so encourages elements on the extras set, $\Delta_e$, to be closer
 to $(\muhat)_{\Delta_e}$ which is nonzero.%

On the other hand, if we instead use the $\ell_\infty$ distance from $(\muhat)_T$, and add it as a constraint, then, at least in certain situations, we can achieve exact recovery with a smaller $n$ than mod-CS. Thus, we study
\bea
\min_\beta \|\beta_{T^c}\|_1 ,   \text{ subject to }  y=A\beta  \text{  and  }  \|\beta_T - \hat{\mu}_T\|_{\infty} \le \rho
\label{rmc}
\eea
and call it {\em reg-mod-BP}.
We see from simulations, that {\em whenever one or more of the inequality constraints are active at $x$, i.e. $|x_i - \hat{\mu}_i | = \rho$ for some $i \in T$, (\ref{rmc}) does achieve exact recovery with fewer measurements than mod-CS}. We use this observation to derive a better exact recovery result below\footnote{One can also try to constrain the $\ell_2$ distance instead of the $\ell_\infty$ distance. When the $\ell_2$ constraint is active, one should again need a smaller $n$ for exact recovery. When we check this via simulations, this does happen, but since it is at most one active constraint, the reduction in $n$ required is small compared to what is achieved by (\ref{rmc}) and hence we do not study this further.}.




\Section{Exact Reconstruction Conditions}
In this section, we obtain exact reconstruction conditions for reg-mod-BP by exploiting the above fact. We give the result and discuss its implications below in Sec III-A. The key lemmas leading to its proof are given in Sec. III-B and the proof outline in Sec. III-C.%

\Subsection{Exact Reconstruction Result}
Let us begin by defining the two types of active sets (set of indices for which the inequality constraint is active), $T_{\text{a+}}$ and $T_{\text{a-}}$, and the inactive set, $T_{\text{in}}$, as follows.
\bea
T_{\text{a+}} \sdefn \{i \in T: x_i - \hat{\mu}_i = \rho \}, \nn \\
T_{\text{a-}} \sdefn \{i \in T: x_i - \hat{\mu}_i = -\rho \}, \nn \\
T_{\text{in}} \sdefn \{i \in T: |x_i - \hat{\mu}_i| < \rho \} \label{Tsplit}
\eea
In the result below,  we try to find the sets $T_{\text{a+g}} \subseteq T_{\text{a+}}$ and $T_{\text{a-g}} \subseteq T_{\text{a-}}$ so that $|T_{\text{a+g}}|+|T_{\text{a-g}}|$ is maximized while $T_{\text{a+g}}$ and $T_{\text{a-g}}$ satisfy certain constraints. We call these the ``good" sets. We define the ``bad" subset of $T$, as $T_b:=T \setminus (T_{\text{a+g}} \cup T_{\text{a-g}})$. As we will see, the smaller the size of this bad set, the weaker are our exact recovery conditions.

\begin{theorem}[Exact Recovery Conditions]
Consider recovering a sparse vector, $x$, with support $N$, from $y:= Ax$ by solving (\ref{rmc}).  The support estimate, $T$, and the misses and extras in it, $\Delta$, $\Delta_e$, satisfy (\ref{defDeltas}). The signal estimate, $\muhat$, satisfies (\ref{feasiblecons}), i.e. $\|x_{T}-\hat{\mu}_{T}\|_{\infty} \le \rho$.
Define the sizes of the sets $T$ and $\Delta$ as
\bea
k:= |T|, \ u:=|\Delta|.
\label{defsizes}
\eea
The true $x$ is the unique minimizer of (\ref{rmc}) if
\begin{enumerate}
\item $\delta_{k+u} < 1, \ \ \delta_{2u} + \delta_{k} + \theta_{k,2u}^2 < 1$,   and
\label{cond1}

\item $a_{k}(2u,u) + a_{\stb}(u,u) < 1$ where 
    \label{cond2}
\bea
T_b \sdefn T \setminus (T_{\text{a+g}} \cup T_{\text{a-g}}), \ \ \text{and} \nn \\
k_b \sdefn |T_b| \nn \\
 \{T_{\text{a+g}},T_{\text{a-g}}\} \se \arg \max_{\tilde{T}_{\text{a+g}}, \tilde{T}_{\text{a-g}} }  (|\tilde{T}_{\text{a+g}}|+|\tilde{T}_{\text{a-g}}|)  \text{ subject to }  \label{cond3} \nn \\
&&  \tilde{T}_{\text{a+g}} \subseteq T_{\text{a+}}, \  \ \tilde{T}_{\text{a-g}} \subseteq T_{\text{a-}}, \label{condset} \nn \\
 && {A_i}'w  > 0 \ \  \forall \ i \in \tilde{T}_{\text{a+g}}, \ \text{and} \ {A_i}'w  < 0 \ \ \forall \ i \in \tilde{T}_{\text{a-g}}, \ \ \ \ \ \ \  \label{c12}  \\
&&  \text{where} \nn \\
&& w \defn M(\tilde{T}_b) A_{\Delta} ({A_{\Delta}}'M(\tilde{T}_b) A_{\Delta})^{-1}\text{sgn}(x_{\Delta}), \nn \\
&& \tilde{T}_b \defn T \setminus (\tilde{T}_{\text{a+g}} \cup \tilde{T}_{\text{a-g}}),  \nn
\eea
$M(S)$ is specified in (\ref{defM}), $a_{k}(s,\Sp)$ is defined in (\ref{def_a}), and the sets $T_{\text{a+}}$, $T_{\text{a-}}$ are defined in (\ref{Tsplit}). 
$\blacksquare$
\end{enumerate}
\label{thm1}
\end{theorem}
%

Notice that $a_k(s,\Sp)$ is a non-decreasing function of $k$. Since $k_b=k - |T_{\text{a+g}}|-|T_{\text{a-g}}|$, thus, finding the largest possible sets $T_{\text{a+g}}$ and $T_{\text{a-g}}$ ensures that the condition $a_{k}(2u,u) + a_{\stb}(u,u) < 1$ is the weakest. The reason for defining $T_{\text{a+g}}$ and $T_{\text{a-g}}$ in the above fashion will become clear in the proof of Lemma \ref{wbnd_iter0}.%

Notice also that the first condition of the above result ensures that $\delta_k < 1$. Since $|\tilde{T}_b| \le k$, thus, ${A_{\tilde{T}_b}}'{A_{\tilde{T}_b}}$ is positive definite and thus invertible. Thus $M(\tilde{T}_b)$ is always well defined. The first condition also ensures that $a_k(2u,u) > 0$. Since $k_b \le k$, and since $\delta_s$ and $\theta_{s_1,s_2}$ are non-decreasing functions of $s, s_1, s_2$, it also ensures that $a_{k_b}(u,u) > 0$. 

\begin{remark}[Applicability]
A practical case where some of the inequality constraints will be active with nonzero probability is when dealing with quantized signals and quantized signal estimates. If the range of values that the signal estimate can take given the signal (or vice versa) is known, the smallest choice of $\rho$ is easily computed. We show some examples in Sec. IV. In general, even if just the range of values both can take is known, we can compute $\rho$.
The fewer the number values that $x_i-\muhat_i$ can take, the larger will be the expected size of the active set, $T_a: = {T}_{\text{a+}} \cup {T}_{\text{a-}}$. Also, the condition (\ref{c12}) will hold for non-empty $T_g:= T_{\text{a+g}} \cup T_{\text{a-g}}$ with nonzero probability. 
\\ 
Some real applications where quantized signals and signal estimates occur are recursive CS based video compression \cite{CSvideosampling,multiscaleCSvideo} (the original video itself is quantized) or in recursive projected CS (ReProCS) \cite{rrpcp_arxiv,rrpcp_isit} based moving or deforming foreground objects' extraction (e.g. a person moving towards a camera) from very large but correlated noise (e.g. very similar looking but slowly changing backgrounds), particularly when the videos are coarsely quantized (low bit rate). A common example where low bit rate videos occur is mobile telephony applications. In any of these applications, if we know a bound on the maximum change of the sparse signal's value from one time instant to the next, that can serve as $\rho$.
%
\end{remark}

\begin{remark}[Comparison with BP, mod-CS, other results]
The worst case for Theorem 1 is when both the sets ${T}_{\text{a+g}}$ and ${T}_{\text{a-g}}$ are empty either because no constraint is active ($T_{\text{a+}}$ and $T_{\text{a-}}$ are both empty) or because (\ref{c12}) does not hold for any pair of subsets of $T_{\text{a+}}$ and $T_{\text{a-}}$. In this case, we  have $k_b=k$ and so the required sufficient conditions are the same as those of mod-CS \cite[Theorem 1]{modcsjournal}. A small extra requirement is that $x$ satisfies (\ref{feasiblecons}). Thus, in the worst case, Theorem 1 holds under the same conditions on $A$  (needs the same number of measurements) as mod-CS \cite{modcsjournal}. In \cite{modcsjournal}, we have already argued that the mod-CS result holds under weaker conditions than the results for BP \cite{decodinglp,candes_rip} as long as the size of the support errors, $|\Delta|, |\Delta_e|$, are small compared to the support size, $|N|$, and hence the same can be said about Theorem 1. For example, we argued that when $|\Delta| = |\Delta_e| = 0.02|N|$ (numbers taken from a recursive dynamic MRI application), the mod-CS conditions are weaker than those of BP. Small $|\Delta|,|\Delta_e|$ is a valid assumption in recursive recovery applications like recursive dynamic MRI, recursive CS based video compression, or ReProCS based foreground extraction from large but correlated background noise.
\\
Moreover, if some inequality constraints are active and (\ref{c12}) holds, as in case of quantized signals and signal estimates, Theorem 1 holds under weaker conditions on $A$ than the mod-CS result.
\\
As noted by an anonymous reviewer, our exact recovery conditions require knowledge of $x$. However this is an issue with many results in sparse recovery, e.g. \cite{justrelax}, and especially those that use more prior knowledge, e.g. \cite{modelCS}.
\end{remark}

\begin{remark}[Small reconstruction error]
The reconstruction error of reg-mod-BP is significantly smaller than that of mod-CS, weighted $\ell_1$ or BP, {\em even when none of the constraints is active}, as long as $\rho$ is small (see Table \ref{not_quantized}). On the other hand, the exact recovery conditions {\em do not} depend on the value of $\rho$, but only on the size of the good subsets of the active sets. This is also observed in our simulations. In Table \ref{not_quantized}, we show results for $\rho = 0.1$. Even when we tried $\rho=0.5$, the exact reconstruction probability or the smallest $n$ needed for exact reconstruction remained the same, but the reconstruction error increased.
\end{remark}

\begin{remark}[Computation complexity]
Finding the best $T_{\text{a+g}}$ and $T_{\text{a-g}}$ requires that one check all possible subsets of $T_{\text{a+}}$ and $T_{\text{a-}}$ and find the pair with the largest sum of sizes that satisfies (\ref{c12}). To do this, one would start with $\tilde{T}_{\text{a+g}} = T_{\text{a+}}$,  $\tilde{T}_{\text{a-g}} = T_{\text{a-}}$; compute $\tilde{T}_b$ and $w$ and check if (\ref{c12}) holds; if it does not, remove one element from $\tilde{T}_{\text{a+g}}$ and then check (\ref{c12}); then remove an element from  $\tilde{T}_{\text{a-g}}$ and check (\ref{c12}); keep doing this until one finds a pair for which (\ref{c12}) holds. In the worst case, one will need to check (\ref{c12}) $2^{|T_{\text{a+}}| + |T_{\text{a-}}|}$ times. However,  the complexity of computing the RIC $\delta_{|T|}$ or any of the ROC's is anyway exponential in $|T|$ and $|T| \ge |T_{\text{a+}}| + |T_{\text{a-}}|$. In summary, computing the conditions of Theorem 1 has complexity that is exponential in the support size, but the same is true for all sparse recovery results that use the RIC. We should mention though that, for certain random matrices, e.g. random Gaussian, there are results that upper bound the RIC values with high probability, e.g. see \cite{decodinglp}. However, the resulting bounds are usually quite loose.
\end{remark} %

\Subsection{Proof of Theorem 1: Key Lemmas}
Our overall proof strategy is similar to that of \cite{decodinglp} for BP and of \cite{modcsjournal} for mod-CS. We first find a set of sufficient conditions on an $n \times 1$ vector, $w$, that help ensure that $x$ is the unique minimizer of (\ref{rmc}). This is done in  Lemma \ref{wcond}. Next, we find sufficient conditions that the measurement matrix $A$ should satisfy so that one such $w$ can be found. This is done in an iterative fashion in the theorem's proof. The proof uses Lemma \ref{wbnd_iter0} at the zeroth iteration, followed by applications of Lemma \ref{wbnd} at later iterations.

To obtain the sufficient conditions on $w$, as suggested in \cite{decodinglp}, we first write out the Karush-Kuhn-Tucker (KKT) conditions for $x$ to be {\em a} minimizer of (\ref{rmc}) \cite[Chapter 5]{boyd}. By strengthening these a little, we get a set of {\em sufficient} conditions for $x$ to be {\em the unique} minimizer.
The necessary conditions for $x$ to be a minimizer are: there exists an $n \times 1$, vector $w$ (Lagrange multiplier for the constraints in $y = A x$),  a $|T_{\text{a+}}| \times 1$ vector, $\lambda_1$, and a $|T_{\text{a-}}| \times 1$ vector, $\lambda_2$, such that (s.t.)
\begin{enumerate}
\item every element of $\lambda_1$ and $\lambda_2$ is non-negative, i.e. $\lambda_1 \succeq 0$ and  $\lambda_2 \succeq 0$,

\item ${A_{T_{\text{in}}}} ' w =0$, ${A_{T_{\text{a+}}}}' w = \lambda_1$, ${A_{T_{\text{a-}}}}' w= - \lambda_2$,  ${A_\Delta} ' w = \text{sgn}(x_\Delta)$, and  $\|{A_{(T \cup \Delta)^c}}'w\|_\infty \le 1$.
\end{enumerate}
As we will see in the proof of Lemma \ref{wcond}, strengthening $\|{A_{(T \cup \Delta)^c}}'w\|_\infty \le 1$ to $\|{A_{(T \cup \Delta)^c}}'w\|_\infty < 1$, keeping the other conditions the same, and requiring that $\delta_{k+u}<1$ gives us a set of {\em sufficient} conditions. 

\begin{lemma}
Let $x$ be as defined in Theorem 1.  $x$ is the unique minimizer of (\ref{rmc}) if $\delta_{k+u}<1$ and if we can find  an $n \times 1$ vector, $w$, s.t.
\begin{enumerate}
\item  ${A_{T_{\text{in}}}} ' w =0$, ${A_{T_{\text{a+}}}}' w \succeq 0$, ${A_{T_{\text{a-}}}}' w  \preceq 0$,

\item ${A_{\Delta}}'w =\text{sgn}(x_\Delta)$,

\item $|{A_j}'w|<1$  for all $j \notin T\cup \Delta$.
\end{enumerate} \label{wcond}
Recall that $T_{\text{a+}}$, $T_{\text{a-}}$ and $T_{\text{in}}$ are defined in (\ref{Tsplit}) and $k,u$ in Theorem 1.  $\blacksquare$
\end{lemma}
{\em Proof: } The proof is given in Appendix \ref{prooflemma1}.


Notice that the first condition is weaker than that of Lemma 1 of mod-CS \cite{modcsjournal} (which requires ${A_{T}}'w =0$), while the other two are the same.
Next, we try to obtain sufficient conditions on the measurement matrix, $A$ (on its RIC's and ROC's) to ensure that such a $w$ can be found. This is done by using Lemmas \ref{wbnd_iter0} and \ref{wbnd} given below. 
Lemma \ref{wbnd_iter0} helps ensure that the first two conditions of Lemma \ref{wcond} hold and provides the starting point for ensuring that the third condition also holds. Then, Lemma \ref{wbnd} applied iteratively helps ensure that the third condition also holds. 

\begin{lemma} 
Assume that $k+u \le m$.
Let $\Sp$ be such that $\st+\sd+\Sp \le m$. If $\delta_{\sd} + \delta_{k_b} + \theta_{k_b,\sd}^2 < 1$, then there exists an $n \times 1$ vector $\tw$ and an ``exceptional" set, $E$, disjoint with $T \cup \Delta$, s.t.  %
\ben
\item  ${A_{T_{b}}}' \tw = 0$,  ${A_{T_{\text{a+g}}}}' \tw \succ 0$, ${A_{T_{\text{a-g}}}}' \tw \prec 0$,
\item ${A_{\Delta}}' \tw = \text{sgn}(x_\Delta)$,

\label{cond3}
\item $|E|  <  \Sp$, $\|{A_E}'\tw\|_2 \le a_{\stb}(\sd,\Sp) \sqrt{\sd}$, $|{A_j}'\tw| \le \frac{a_{\stb}(\sd,\Sp)}{\sqrt{\Sp}} \sqrt{\sd} \  \ \forall j \notin T \cup \Delta \cup E $,

\label{cond4}
\item $\|\tw\|_2 \le K_{\stb}(\sd) \sqrt{\sd}$.
\een
Recall that  $a_{\st}(s,\Sp)$, $K_k(s)$ are defined in (\ref{def_a}), (\ref{def_K}) and $T_{\text{a+g}}$, $T_{\text{a-g}}$, $T_b$, $k_b$, $k$ and $u$ in Theorem \ref{thm1}.  $\blacksquare$
\label{wbnd_iter0}
\end{lemma}

Notice that because we have assumed that $\delta_{\sd} + \delta_{k_b} + \theta_{k_b,\sd}^2 < 1$, $a_{\stb}(\sd,\Sp)$ and $K_{\stb}(\sd)$ are positive.
We call the set $E$ an ``exceptional" set, because except on the set $E \subseteq (T \cup \Delta)^c$, everywhere else on $(T \cup \Delta)^c$, $|{A_j}'\tw|$ is bounded. This notion is taken from \cite{decodinglp}.
Notice that the first two conditions of the above lemma are one way to satisfy the first two conditions of Lemma \ref{wcond} since $T_b = T_{\text{in}} \cup (T_{\text{a+}} \setminus T_{\text{a+g}}) \cup (T_{\text{a-}} \setminus T_{\text{a-g}})$.

{\em Proof: } The proof is given in Appendix \ref{prooflemma3}. We let $\tw=M(T_b) A_{\Delta} ({A_{\Delta}}'M(T_b) A_{\Delta})^{-1} \text{sgn}(x_{\Delta})$. Since the good sets $T_{\text{a+g}}$, $T_{\text{a-g}}$ are appropriately defined (see (\ref{c12})), the first two conditions hold. The rest of the proof bounds $\|\tw\|_2$, and finds the set $E \subseteq (T \cup \Delta)^c$ of size $|E| < \Sp$ so that $|{A_j}'\tw|$ is bounded for all $i \notin T \cup \Delta \cup E$ and also $\|{A_E}'\tw\|_2$ is bounded.

\begin{lemma} [Lemma 2 of \cite{modcsjournal}] 
Assume that $k \le m$.
Let $s$, $\Sp$ be such that $\st+s+\Sp \le m$. Assume that $\delta_{s} + \delta_{\st} + \theta_{\st,s}^2 < 1$. Let $T_d$ be a set that is disjoint with $T$, of size $|T_d| \le s$ and let $c$ be a $|T_d| \times 1$ vector.  Then there exists an $n \times 1$ vector, $\tw$, and a set, $E$, disjoint with $T \cup T_d$, s.t. 
(i) ${A_T}'\tw = 0$,
(ii) ${A_{T_d}}'\tw = c$,
(iii) $|E|  <  \Sp$, $\|{A_E}'\tw\|_2 \le a_{\st}(s,\Sp) \|c\|_2$, $|{A_j}'\tw| \le \frac{a_{\st}(s,\Sp)}{\sqrt{\Sp}} \|c\|_2, \  \forall j \notin T \cup T_d \cup E$,  and
(iv) $\|\tw\|_2 \le K_{\st}(s) \|c\|_2$.
\\ Recall that $a_{\st}(s,\Sp)$, $K_k(s)$ are defined in (\ref{def_a}), (\ref{def_K}), and $k,u$ in Theorem 1.   $\blacksquare$
\label{wbnd}
\end{lemma}

{\em Proof: } The proof of Lemma \ref{wbnd} is given in \cite{modcsjournal} and also in Appendix C of \cite{regmodBParxiv}.

Notice that because we have assumed that $\delta_{s} + \delta_{\st} + \theta_{\st,s}^2 < 1$, $a_{\st}(s,\Sp)$ and $K_{\st}(s)$ are positive.

\Subsection{Proof Outline of Theorem \ref{thm1}}
The proof is very similar to that of \cite{modcsjournal}. Hence we give only the outline here. The complete proof is in \cite{regmodBParxiv}. At iteration zero, we apply Lemma \ref{wbnd_iter0} with $\Sp \equiv \sd$, to get a $w_1$ and an exceptional set $T_{d,1}$, disjoint with $T \cup \Delta$, of size less than $\sd$. Lemma \ref{wbnd_iter0} can be applied because  $k_b \le k$ and condition \ref{cond1} of the theorem holds. At iteration $\iter > 0$, we apply Lemma \ref{wbnd} with $T_{d} \equiv \Delta \cup T_{d,\iter}$ (so that $s \equiv 2\sd$), $c_\Delta \equiv 0$, $c_{T_d} \equiv {A_{T_d}}' w_\iter$ and $\Sp \equiv \sd$ to get a $w_{r+1}$ and an exceptional set $T_{d,\iter+1}$ disjoint with $T \cup \Delta \cup T_{d,\iter}$ of size less than $\sd$. Lemma \ref{wbnd} can be applied because condition \ref{cond1} of the theorem holds. Define $w \defn \sum_{\iter=1}^\infty (-1)^{\iter-1} w_\iter$. We then argue that if  condition \ref{cond2} of the theorem holds, $w$ is well-defined and satisfies the conditions of Lemma \ref{wcond}. Applying Lemma \ref{wcond}, the result follows.
%
\input{Final_Revision_expts}

\Section{Conclusions} 
In this work, we obtained sufficient exact recovery conditions for reg-mod-BP, (\ref{rmc}), and discussed their implications. Our main conclusion is that if some of the inequality constraints are active and if even a subset of the set of active constraints satisfies certain conditions (given in (\ref{c12})), then reg-mod-BP achieves exact recovery under weaker conditions than what mod-CS needs. A practical situation where this would happen is when both the signal and its estimate are quantized. In other cases, the conditions are only as weak as those for mod-CS. In either case they are much weaker than those for BP as long as $T$ is a good support estimate.
From simulations, we see that even without any active constraints, the reg-mod-BP reconstruction error is much lower than that of mod-CS or weighted $\ell_1$.


\appendix

\Subsection{Proof of Lemma \ref{wcond}} \label{prooflemma1}
 Denote a minimizer of (\ref{rmc}) by $\beta$. Since $y=Ax$ and $x$ satisfies (\ref{feasiblecons}), $x$ is feasible for (\ref{rmc}). Thus,
\bea
\|\beta_{T^c}\|_1 \le \|x_{T^c}\|_1=\|x_\Delta\|_{1}
\label{eq0}
\eea
Next, we use the conditions on $w$ given in Lemma \ref{wcond} and the fact that $x$ is supported on $N \subseteq T \cup \Delta$  to show that $\|\beta_{T^c}\|_1 \ge \|x_{T^c}\|_1$ and hence $\|x_{T^c}\|_1 = \|\beta_{T^c}\|_1$. Notice that
\bea
\label{eq1}
\|\beta_{T^c}\|_1 \se \sum_{j\in \Delta}|x_j+\beta_j-x_j|+\sum_{j\notin T\cup \Delta}|\beta_j| \ge \sum_{j\in \Delta}|x_j+\beta_j-x_j|+\sum_{j \notin T\cup \Delta} w'A_j\beta_j   \\
\label{eq2}
\sge \sum_{j\in \Delta}\text{sgn}(x_j)(x_j+(\beta_j-x_j))+\sum_{j \notin T\cup \Delta} w'A_j (\beta_j-x_j) \\
\label{eq4}
\se \|x_{\Delta}\|_1 +\sum_{j \notin T} w'A_j (\beta_j-x_j)
= \|x_{\Delta}\|_1 + w'(A\beta-Ax)-\sum_{j\in T}w'A_j(\beta_j-x_j)  \\
\label{eq5}
\se \|x_{\Delta}\|_1-\sum_{j\in T}w'A_j(\beta_j-\hat{\mu}_j+\hat{\mu}_j-x_j) \\
\label{eq6}
\se \|x_{\Delta}\|_1-\sum_{j\in T_{\text{a+}}}w'A_j(\beta_j-\hat{\mu}_j-\rho) -\sum_{j\in T_{\text{a-}}}w'A_j(\beta_j-\hat{\mu}_j+\rho)  \\
\label{eq7}
\sge \|x_{\Delta}\|_1 = \|x_{T^c}\|_1
\eea
%
In the above, the inequality in (\ref{eq1}) follows because $w'A_j \le |w'A_j| < 1$ for $j \notin T \cup \Delta$ and because $|\beta_j| \ge \beta_j$. Inequality (\ref{eq2}) uses the fact that $|z| \ge \sgn(b) z$ for any two scalars $z$ and $b$ and that $x_j=0$ for $j \notin T \cup \Delta$. In (\ref{eq4}), the first equality uses $\sgn(x_j)x_j = |x_j|$ and $w'A_j = \sgn(x_j)$ for $j \in \Delta$. The second equality just rewrites the second term in a different form. In (\ref{eq5}), we use the fact that $A \beta = A x = y$ (since both $\beta$ and $x$ are feasible) to eliminate $w'(A\beta-Ax)$. Equation (\ref{eq6}) uses $w'A_j =0$ for $j \in T_{\text{in}}$ and the definitions of $T_{\text{a+}}$ and $T_{\text{a-}}$ given in (\ref{Tsplit}). Finally, (\ref{eq7}) follows because  $-\sum_{j \in T_{\text{a+}}} w'A_j(\beta_j-\hat{\mu}_j-\rho) -\sum_{j\in T_{\text{a-}}}w'A_j(\beta_j-\hat{\mu}_j+\rho) \ge 0$. This holds since $-\rho \le \beta_j-\hat{\mu}_j \le \rho$ for all $j \in T$; $w'A_j \ge 0$ for $j \in T_{\text{a+}}$; and $w'A_j \le 0$ for $j \in T_{\text{a-}}$.

Both inequalities (\ref{eq0}) and (\ref{eq1})-(\ref{eq7}) can hold only when $\|\beta_{T^c}\|_1=\|x_{T^c}\|_1$, i.e. all the inequalities in (\ref{eq1})-(\ref{eq7}) hold with equality. Consider the inequality in (\ref{eq1}). Since $|w'A_j|<1$ for $j\notin T\cup \Delta$, this holds with equality only if $\beta_j=0$ for all $j\notin T\cup \Delta$. Since $A\beta=y=Ax$ and since both $\beta$ and $x$ are supported on $T\cup \Delta$ (or on its subset), $A_{T\cup \Delta}(\beta_{T\cup \Delta}-x_{T\cup \Delta})=0$. Since $\delta_{k+u}<1$, $A_{T\cup \Delta}$ has full rank. Therefore, this means that $\beta_{T\cup \Delta}=x_{T\cup \Delta}$. Thus, we can conclude that $\beta=x$, i.e., $x$ is the unique minimizer.

\Subsection{Proof of Lemma \ref{wbnd_iter0}}\label{prooflemma3}
This proof uses the following simple facts. Let $\lambda_{\min}(M)$, $\lambda_{\max}(M)$ denote the minimum and maximum eigenvalues of a matrix $M$.
(i) For positive semi-definite matrices, $M$, $Q$, $\|M\| = \lambda_{\max}(M)$; $\|M Q\| \le \|M\| \|Q\|$; $\lambda_{\min}(M-Q) \ge \lambda_{\min}(M) - \lambda_{\max}(Q)$; and for a positive definite matrix, $M$, $\|M^{-1} \| = 1/\lambda_{\min}(M)$;
(ii) for any matrices, $B$, $C$, $\|B-C\| \le \|B\| + \|C\|$;
(iii) for disjoint sets $T_1, T_2$, $\|{A_{T_1}}'A_{T_2}\| \le \theta_{|T_1|,|T_2|}$ \cite[equation (3)]{modcsjournal};
(iv) $1-\delta_{|T_1|} \le \lambda_{\min}({A_{T_1}}'A_{T_1}) \le \lambda_{\max}({A_{T_1}}'A_{T_1}) \le 1+\delta_{|T_1|}$ \cite{decodinglp};
(v) $M(T_b)$ is a projection matrix and so $M(T_b)M(T_b)' = M(T_b)$ and $\|M(T_b)\|=1$; 
(vi) $\|\sgn(x_{\Delta})\|_2 = \sqrt{u}$.

The lemma assumes that $\delta_{\sd} + \delta_{\stb} + \theta_{\stb,\sd}^2 < 1$. This implies that (a) $\delta_{\sd} < 1$ and so ${A_\Delta}'A_\Delta$ is positive definite and so $u \le n$; (b) $\delta_{\stb} < 1$ and so ${A_{T_b}}'{A_{T_b}}$ is positive definite and $M(T_b)$ is well-defined; and (c) as we show next, ${A_\Delta}' M(T_b) A_\Delta$ is positive definite and hence full rank.  Since ${A_\Delta}' M(T_b) A_\Delta = {A_\Delta}' A_\Delta - {A_\Delta}' {A_{T_b}} ({A_{T_b}}'{A_{T_b}})^{-1} {A_{T_b}}' A_\Delta$ is a difference of two positive semi-definite matrices, thus,
\bea
\lambda_{\min}({A_{\Delta}}'M(T_b) A_{\Delta} ) \ge \lambda_{\min}({A_\Delta}' A_\Delta) - \lambda_{\max}({A_\Delta}' {A_{T_b}} ({A_{T_b}}'{A_{T_b}})^{-1} {A_{T_b}}' A_\Delta) \ge (1-\delta_u) - \frac{\theta_{k_b,u}^2}{1-\delta_{k_b}} > 0  \ \ \ \ \
\label{bnd_lmin}
\eea
Thus, ${A_{\Delta}}'M(T_b) A_{\Delta}$ is positive definite.
The first inequality in (\ref{bnd_lmin}) follows from fact (i). The second one follows because $\lambda_{\min}({A_\Delta}' A_\Delta)  \ge (1-\delta_u)$ (using fact (iv)); $\lambda_{\max}({A_\Delta}' {A_{T_b}} ({A_{T_b}}'{A_{T_b}})^{-1} {A_{T_b}}' A_\Delta) = \|{A_\Delta}' {A_{T_b}} ({A_{T_b}}'{A_{T_b}})^{-1} {A_{T_b}}' A_\Delta\| \le \|{A_\Delta}' {A_{T_b}} \| \ \|({A_{T_b}}'{A_{T_b}})^{-1}\| \ \|{A_{T_b}}' A_\Delta\|$ (using fact (i)); $\|{A_\Delta}' {A_{T_b}} \| = \|{A_{T_b}}'A_\Delta \| \le \theta_{k_b,u}$ (using fact (iii)); and $\|({A_{T_b}}'{A_{T_b}})^{-1}\| = \frac{1}{ \lambda_{\min}({A_{T_b}}'{A_{T_b}}) } \le \frac{1}{1-\delta_{k_b}}$ (since ${A_{T_b}}'{A_{T_b}}$ is positive definite, this follows using fact (i) and fact (iv)).
The third inequality of (\ref{bnd_lmin}) follows because $(1-\delta_u) - \frac{\theta_{k_b,u}^2}{1-\delta_{k_b}} = \frac{1-\delta_u -\delta_{k_b} + \delta_u \delta_{k_b} - \theta_{k_b,u}^2}{1-\delta_{k_b}} > 0$. Both the numerator and the denominator are positive because we have assumed that $\delta_{\sd} + \delta_{\stb} + \theta_{\stb,\sd}^2 < 1$.


Using fact (v), ${A_\Delta}' M(T_b) A_\Delta = {A_\Delta}' M(T_b) M(T_b)' A_\Delta$. Thus, using the above, ${A_\Delta}' M(T_b) M(T_b)' A_\Delta$ is positive definite and hence has full rank $u$. Thus, the $u \times n$ fat matrix, ${A_\Delta}' M(T_b)$ has full rank, $u$.

To prove the lemma, we first try to construct an $n \times 1$ vector, $\tw$, that satisfies the first two conditions of the lemma. Then, we show that we can find an exceptional set $E$ so that the constructed $\tw$ and $E$ satisfy all the required conditions.
Any $\tw$ that satisfies ${A_{T_b}}' \tw = 0$ lies in the null space of ${A_{T_b}}'$ and hence is of the form $\tw = M(T_b) \gamma$. To satisfy the second condition, we need a $\gamma$ that satisfies ${A_\Delta}'M(T_b) \gamma = \sgn(x_\Delta)$. As shown above, ${A_\Delta}'M(T_b)$ is full rank and so this system of equations has a solution (in fact has infinitely many solutions). We can compute the minimum $\ell_2$ norm solution in closed form as $\gamma = M(T_b)'{A_\Delta}({A_\Delta}'M(T_b) M(T_b)' {A_\Delta})^{-1} \sgn(x_\Delta)$. Since $M(T_b)M(T_b)' = M(T_b)$, $\tw = M(T_b) \gamma$ can be rewritten as
\bea
\tw = M(T_b) A_{\Delta} ({A_{\Delta}}'M(T_b) A_{\Delta})^{-1} \text{sgn}(x_{\Delta})
\label{deftw}
\eea
Using the definition of $T_{\text{a+g}}$, $T_{\text{a-g}}$ given in (\ref{c12}) in Theorem 1, we can see that $\tw$ satisfies the first two conditions of the lemma. Recall that ${A_i}'w > 0$ for all $i \in T_{\text{a+g}}$ is equivalent to ${A_{T_{\text{a+g}}}}'w \succ 0$, and similarly, ${A_i}'w < 0$ for all $i \in T_{\text{a-g}}$ is equivalent to ${A_{T_{\text{a-g}}}}'w \prec 0$.

The rest of the proof is similar to that of \cite[Lemma 2]{modcsjournal}. Consider any set $\Tdp$ disjoint with $T \cup \Delta$ of size $|\Tdp| \le \Sp$. Then,%
\bea
\|{A_{\Tdp}}'\tw \|_2
\sle \|{A_{\Tdp}}' M(T_b) A_{\Delta}\| \ \|({A_{\Delta}}'M(T_b) A_{\Delta} )^{-1}\| \ \|\text{sgn}(x_{\Delta})\|_2  \nn \\
\sle (\theta_{\Sp,u} + \frac{\theta_{\Sp,k_b} \theta_{u,k_b}}{1-\delta_{k_b}}) \frac{1}{1-\delta_u - \frac{\theta_{u,k_b}^2}{1-\delta_{k_b}}} \sqrt{\sd} = a_{\stb}(\sd,\Sp) \sqrt{\sd}
\label{def_au}
\eea
Notice that $a_{\stb}(\sd,\Sp)$ is positive because we have assumed that $\delta_{\sd} + \delta_{\stb} + \theta_{\stb,\sd}^2 < 1$. The bound in (\ref{def_au}) follows using the simple facts given in the beginning. We obtain (\ref{def_au}) as follows. Consider the first term $\|{A_{\Tdp}}' M(T_b) A_{\Delta}\|$.  Using the definition of $M(T_b)$ and fact (ii), $\|{A_{\Tdp}}'M(T_b) A_\Delta\| \le \|{A_{\Tdp}}'A_\Delta \| + \|{A_{\Tdp}}'{A_{T_b}} ({A_{T_b}}'{A_{T_b}})^{-1} {A_{T_b}}' A_\Delta \|$. Using fact (iii), $\|{A_{\Tdp}}'A_\Delta \| \le \theta_{\Sp,u}$,
 $\|{A_{\Tdp}}'{A_{T_b}}\| \le \theta_{\Sp,k_b}$ and $\| {A_{T_b}}' A_\Delta \| \le \theta_{u,k_b}$. Since ${A_{T_b}}'{A_{T_b}}$ is positive definite, using fact (i) and fact (iv), $\|({A_{T_b}}'{A_{T_b}})^{-1}\| = \frac{1}{ \lambda_{\min}({A_{T_b}}'{A_{T_b}}) } \le \frac{1}{1-\delta_{k_b}}$. Thus, we get $ \|{A_{\Tdp}}' M(T_b) A_{\Delta}\| \le (\theta_{\Sp,u} + \frac{\theta_{\Sp,k_b} \theta_{u,k_b}}{1-\delta_{k_b}})$.
Consider the second term $\|({A_{\Delta}}'M(T_b) A_{\Delta} )^{-1}\|$. Since ${A_{\Delta}}'M(T_b) A_{\Delta}$ is positive definite, using fact (i) and (\ref{bnd_lmin}), $\|({A_{\Delta}}'M(T_b) A_{\Delta} )^{-1}\| = \frac{1}{\lambda_{\min}({A_{\Delta}}'M(T_b) A_{\Delta} )} \le \frac{1}{(1-\delta_u) - \frac{\theta_{u,k_b}^2}{1-\delta_{k_b}}}$. Using fact (vi), the third term, $\|\text{sgn}(x_{\Delta})\|_2= \sqrt{\sd}$.




Define the set, $E$, as $E := \{ j \in (T \cup \Delta)^c : |{A_j}'\tw| > \frac{a_{\stb}(\sd,\Sp)\sqrt{\sd} }{\sqrt{\Sp}} \}$.  Notice that $|E|$ must obey $|E| < \Sp$ since otherwise we can contradict (\ref{def_au}) by taking $\Tdp \subseteq E$.
Since $|E| < \Sp$ and $E$ is disjoint with $T \cup \Delta$, (\ref{def_au}) holds for $\Tdp \equiv E$, i.e.,
$\|{A_{E}}'\tw \|_2 \le a_{\stb}(\sd,\Sp) \sqrt{\sd}$. Also, by definition of $E$, $|{A_j}'\tw| \le  \frac{a_{\stb}(\sd,\Sp)  \sqrt{\sd}}{\sqrt{\Sp}}, \ \text{for all} \ j \notin T \cup \Delta \cup E$. Thus $\tw$ satisfies the third condition of the lemma.

Finally,  $\|\tw\|_2 \le \| M(T_b)\| \ \|A_{\Delta} \| \ \| ({A_{\Delta}}'M(T_b) A_{\Delta} )^{-1} \| \  \sqrt{\sd} \le  K_{\stb}(\sd) \sqrt{\sd} $. This follows using fact (v);  $\|A_{\Delta} \| \le \sqrt{1 + \delta_{\sd}}$; and fact (i) and (\ref{bnd_lmin}).
Thus, we have found a $\tw$ and $E$ that satisfy all required conditions.%

\linespread{1.1}
\bibliographystyle{IEEEtran}
\bibliography{regmodBPDNjournalbib,regmodCSbib}

\clearpage \input{SecondRound_Revision_SupplemMaterial}

\end{document}

%% file: zcom.tex
\newcommand{\st}{k}
\newcommand{\sd}{u}
\newcommand{\stin}{k_{in}}
\newcommand{\Tin}{T_{in}}
\newcommand{\tw}{\tilde{w}}
\newcommand{\sde}{e}
\newcommand{\sn}{s}
\newcommand{\iter}{r}
\newcommand{\Tdp}{\check{T}_d} 
\newcommand{\sno}{n}
\newcommand{\mno}{m} 

\newcommand{\tty}{\tilde{y}_{t,\text{res}}}
\newcommand{\xhat}{\hat{x}}
\newcommand{\betahat}{\hat{\beta}}
\newcommand{\xhatold}{\hat{x}_{t,\text{old}}}

\newcommand{\Nhat}{\hat{N}}
\newcommand{\Dnum}{D_{num}}
\newcommand{\pss}{p^{**,i}}
\newcommand{\fr}{f_{r}^i}

\newcommand{\A}{{\cal A}}
\newcommand{\Z}{{\cal Z}}
\newcommand{\B}{{\cal B}}
\newcommand{\R}{{\cal R}}
\newcommand{\reg}{{\cal G}}
\newcommand{\const}{\mbox{const}}

\newcommand{\trace}{\mbox{trace}}

\newcommand{\hsim}{{\hspace{0.0cm} \sim  \hspace{0.0cm}}}
\newcommand{\he}{{\hspace{0.0cm} =  \hspace{0.0cm}}}

\newcommand{\vect}[2]{\left[\begin{array}{cccccc}
     #1 \\
     #2
   \end{array}
  \right]
  }

\newcommand{\matr}[2]{ \left[\begin{array}{cc}
     #1 \\
     #2
   \end{array}
  \right]
  }
\newcommand{\vc}[2]{\left[\begin{array}{c}
     #1 \\
     #2
   \end{array}
  \right]
  }

\newcommand{\gdot}{\dot{g}}
\newcommand{\Cdot}{\dot{C}}
\newcommand{\re}{\mathbb{R}}
\newcommand{\n}{{\cal N}}  
\newcommand{\N}{{\overrightarrow{\bf N}}}  
\newcommand{\chat}{\tilde{C}_t}
\newcommand{\chati}{\chat^i}

\newcommand{\cmin}{C^*_{min}}
\newcommand{\twi}{\tilde{w}_t^{(i)}}
\newcommand{\twj}{\tilde{w}_t^{(j)}}
\newcommand{\wi}{{w}_t^{(i)}}
\newcommand{\twio}{\tilde{w}_{t-1}^{(i)}}

\newcommand{\tWi}{\tilde{W}_n^{(m)}}
\newcommand{\tWj}{\tilde{W}_n^{(k)}}
\newcommand{\Wi}{{W}_n^{(m)}}
\newcommand{\tWio}{\tilde{W}_{n-1}^{(m)}}

\newcommand{\ds}{\displaystyle}

\newcommand{\SAR}{S$\!$A$\!$R }
\newcommand{\MAR}{MAR}
\newcommand{\MMRF}{MMRF}
\newcommand{\AR}{A$\!$R }
\newcommand{\GMRF}{G$\!$M$\!$R$\!$F }
\newcommand{\DTM}{D$\!$T$\!$M }
\newcommand{\MSE}{M$\!$S$\!$E }
\newcommand{\RCS}{R$\!$C$\!$S }
\newcommand{\uomega}{\underline{\omega}}
\newcommand{\y}{v}
\newcommand{\x}{w}
\newcommand{\lu}{\mu}
\newcommand{\g}{g}
\newcommand{\s}{{\bf s}}
\newcommand{\bft}{{\bf t}}
\newcommand{\refmap}{{\cal R}}
\newcommand{\totrefl}{{\cal E}}
\newcommand{\beq}{\begin{equation}}
\newcommand{\eeq}{\end{equation}}
\newcommand{\bdm}{\begin{displaymath}}
\newcommand{\edm}{\end{displaymath}}
\newcommand{\hatz}{\hat{z}}
\newcommand{\hatu}{\hat{u}}
\newcommand{\tilz}{\tilde{z}}
\newcommand{\tilu}{\tilde{u}}
\newcommand{\hhatz}{\hat{\hat{z}}}
\newcommand{\hhatu}{\hat{\hat{u}}}
\newcommand{\tilc}{\tilde{C}}
\newcommand{\hatc}{\hat{C}}
\newcommand{\tim}{n}

\newcommand{\ssp}{\renewcommand{\baselinestretch}{1.0}}
\newcommand{\defd}{\mbox{$\stackrel{\mbox{$\triangle$}}{=}$}}
\newcommand{\goes}{\rightarrow}
\newcommand{\tends}{\rightarrow}
\newcommand{\defn}{:=} 
\newcommand{\se}{&=&}
\newcommand{\sdefn}{& \defn  &}
\newcommand{\sle}{& \le &}
\newcommand{\sge}{& \ge &}
\newcommand{\plusminus}{\stackrel{+}{-}}
\newcommand{\Ey}{E_{Y_{1:t}}}
\newcommand{\ey}{E_{Y_{1:t}}}

\newcommand{\equivto}{\mbox{~~~which is equivalent to~~~}}
\newcommand{\nonzero}{i:\pi^n(x^{(i)})>0}
\newcommand{\nonzeroc}{i:c(x^{(i)})>0}

\newcommand{\supn}{\sup_{\phi:\|\phi\|_\infty \le 1}}
\newtheorem{theorem}{Theorem}
\newtheorem{lemma}{Lemma}
\newtheorem{proposition}{Proposition}
\newtheorem{corollary}{Corollary}
\newtheorem{definition}{Definition}
\newtheorem{remark}{Remark}
\newtheorem{example}{Example}
\newtheorem{ass}{Assumption}
\newtheorem{fact}{Fact}
\newtheorem{heuristic}{Heuristic}
\newcommand{\eps}{\epsilon}
\newcommand{\bd}{\begin{definition}}
\newcommand{\ed}{\end{definition}}
\newcommand{\udq}{\underline{D_Q}}
\newcommand{\td}{\tilde{D}}
\newcommand{\epsinv}{\epsilon_{inv}}
\newcommand{\al}{\mathcal{A}}

\newcommand{\bfx} {\bf X}
\newcommand{\bfy} {\bf Y}
\newcommand{\bfz} {\bf Z}
\newcommand{\ddas}{\mbox{${d_1}^2({\bf X})$}}
\newcommand{\ddbs}{\mbox{${d_2}^2({\bfx})$}}
\newcommand{\dda}{\mbox{$d_1(\bfx)$}}
\newcommand{\ddb}{\mbox{$d_2(\bfx)$}}
\newcommand{\xinc}{{\bfx} \in \mbox{$C_1$}}
\newcommand{\eqa}{\stackrel{(a)}{=}}
\newcommand{\eqb}{\stackrel{(b)}{=}}
\newcommand{\eqe}{\stackrel{(e)}{=}}
\newcommand{\leqc}{\stackrel{(c)}{\le}}
\newcommand{\leqd}{\stackrel{(d)}{\le}}

\newcommand{\leqa}{\stackrel{(a)}{\le}}
\newcommand{\leqb}{\stackrel{(b)}{\le}}
\newcommand{\leqe}{\stackrel{(e)}{\le}}
\newcommand{\leqf}{\stackrel{(f)}{\le}}
\newcommand{\leqg}{\stackrel{(g)}{\le}}
\newcommand{\leqh}{\stackrel{(h)}{\le}}
\newcommand{\leqi}{\stackrel{(i)}{\le}}
\newcommand{\leqj}{\stackrel{(j)}{\le}}

\newcommand{\w}{{W^{LDA}}}
\newcommand{\halpha}{\hat{\alpha}}
\newcommand{\hsigma}{\hat{\sigma}}
\newcommand{\slmax}{\sqrt{\lambda_{max}}}
\newcommand{\slmin}{\sqrt{\lambda_{min}}}
\newcommand{\lmax}{\lambda_{max}}
\newcommand{\lmin}{\lambda_{min}}

\newcommand{\da} {\frac{\alpha}{\sigma}}
\newcommand{\chka} {\frac{\check{\alpha}}{\check{\sigma}}}
\newcommand{\sumo}{\sum _{\underline{\omega} \in \Omega}}
\newcommand{\distance}{d\{(\hatz _x, \hatz _y),(\tilz _x, \tilz _y)\}}
\newcommand{\col}{{\rm col}}
\newcommand{\rcs}{\sigma_0}
\newcommand{\CalR}{{\cal R}}
\newcommand{\df}{{\delta p}}
\newcommand{\dq}{{\delta q}}
\newcommand{\dZ}{{\delta Z}}
\newcommand{\pprime}{{\prime\prime}}

\newcommand{\vn}{N}
\newcommand{\diff}{\text{diff}}

\newcommand{\bv}{\begin{vugraph}}
\newcommand{\ev}{\end{vugraph}}
\newcommand{\bi}{\begin{itemize}}
\newcommand{\ei}{\end{itemize}}
\newcommand{\ben}{\begin{enumerate}}
\newcommand{\een}{\end{enumerate}}
\newcommand{\be}{\protect\[}
\newcommand{\ee}{\protect\]}
\newcommand{\bean}{\begin{eqnarray*} }
\newcommand{\eean}{\end{eqnarray*} }
\newcommand{\bea}{\begin{eqnarray} }
\newcommand{\eea}{\end{eqnarray} }
\newcommand{\nn}{\nonumber}
\newcommand{\ba}{\begin{array} }
\newcommand{\ea}{\end{array} }
\newcommand{\ep}{\mbox{\boldmath $\epsilon$}}
\newcommand{\epp}{\mbox{\boldmath $\epsilon '$}}
\newcommand{\Lep}{\mbox{\LARGE $\epsilon_2$}}
\newcommand{\und}{\underline}
\newcommand{\pdif}[2]{\frac{\partial #1}{\partial #2}}
\newcommand{\odif}[2]{\frac{d #1}{d #2}}
\newcommand{\dt}[1]{\pdif{#1}{t}}
\newcommand{\urho}{\underline{\rho}}

\newcommand{\spc}{{\cal S}}
\newcommand{\tspc}{{\cal TS}}

\newcommand{\uv}{\underline{v}}
\newcommand{\us}{\underline{s}}
\newcommand{\uc}{\underline{c}}
\newcommand{\utheta}{\underline{\theta}^*}
\newcommand{\ualpha}{\underline{\alpha^*}}

\newcommand{\uxy}{\underline{x}^*}
\newcommand{\uxyj}{[x^{*}_j,y^{*}_j]}
\newcommand{\arcl}[1]{arclen(#1)}
\newcommand{\one}{{\mathbf{1}}}

\newcommand{\uxyjt}{\uxy_{j,t}}
\newcommand{\E}{\mathbb{E}}

\newcommand{\rhomat}{\left[\begin{array}{c}
                        \rho_3 \ \rho_4 \\
                        \rho_5 \ \rho_6
                        \end{array}
                   \right]}
\newcommand{\deltat}{\tau} 
\newcommand{\deltatt}{\Delta t_1}
\newcommand{\ceil}[1]{\ulcorner #1 \urcorner}

\newcommand{\xxi}{x^{(i)}}
\newcommand{\txi}{\tilde{x}^{(i)}}
\newcommand{\txj}{\tilde{x}^{(j)}}

\newcommand{\mi}[1]{{#1}^{(m,i)}}

%% file: Final_Revision_expts.tex

\Section{Numerical Experiments}
In this section, we show two types of numerical experiments. The first simulates quantized signals and signal estimates. This is the case where some constraints are active with nonzero probability. The good set, $T_g = {T}_{\text{a+g}} \cup {T}_{\text{a-g}}$ is also non empty with nonzero probability. Hence, for a given small enough $n$, reg-mod-BP has significantly higher exact reconstruction probability, $p_{\text{exact}}(n)$, as compared to both mod-CS \cite{modcsjournal} and weighted $\ell_1$ \cite{weightedl1} and much higher than that of BP \cite{BPDN,decodinglp}.  Alternatively, it also requires a significantly reduced $n$ for exact reconstruction with probability one, $n_{\text{exact}}(1)$. In computing $p_{\text{exact}}(n)$ we average over the distribution of $x$, $T$ and $\muhat$, as also in \cite{modcsjournal,decodinglp}. All numbers are computed based on 100 Monte Carlo simulations. To compute $n_{\text{exact}}(1)$, we tried various values of $n$ for each algorithm and computed the smallest $n$ required for exact recovery always (in all 100 simulations).

We also do a second simulation where signal estimates are not quantized.

In the following steps, the notation $z \sim \text{discrete-uniform}(a_1,a_2, \dots a_n)$ means that $z$ is equally likely to be equal to $a_1$, $a_2$, $\dots$ or $a_n$.  We use $\pm a$ as short for $+a,-a$. Also, $z \sim \text{uniform}(a,b)$ generates a scalar uniform random variable in the range $[a,b]$. The notation $x_i \stackrel{iid}{\sim} \text{P}$ for all $i \in S$ means that, for all $i \in S$, each $x_i$ is  identically distributed according to $\text{P}$ and is independent of all the others. 

\begin{table*}[t!]
\center
 \begin{tabular}{|c|c|c|c|c|c|}
\hline
  & {\bf $2K$} & {\bf BP} & {\bf mod-CS} & {\bf weighted $\ell_1$}  & {\bf Reg-mod-BP}  \\
\hline
$p_{\text{exact}}(0.15m)$ & 4 & 0 &0.18 &0.16 & 0.64 \\
\hline
N-RMSE($0.15m$) & 4  & 1.011 & 0.059 & 0.060 &0.029  \\ \hline
$n_{\text{exact}}(1)$ & 4 & 0.39$m$ & 0.21$m$  & 0.21$m$ & 0.18$m$ \\
\hline
\hline
$p_{\text{exact}}(0.15m)$ & 10 & 0 & 0.18 &0.16 & 0.39 \\
\hline
N-RMSE($0.15m$) & 10 & 1.011 & 0.059 & 0.060 & 0.032  \\ \hline
$n_{\text{exact}}(1)$ & 10 & 0.4$m$ & 0.21$m$  & 0.21$m$ & 0.20$m$ \\
\hline
\end{tabular}
\vspace{-0.05in}
\caption{\small{Quantized signals and signal estimates. Recall that $k=|T|=26$. For $2K=4$, the expected sizes of $T_a$, $T_g$ and $T_b$ are $\E[|T_a|]=10.01$, $\E[|T_g|]=5.27$ and $\E[|T_b|]=20.73$. For $2K=10$, $\E[|T_a|]=4.28$, $\E[|T_g|]=2.3$ and $\E[|T_b|]=23.7$. }}
\vspace{-0.05in}
\label{quantized_1}
\end{table*} 

\begin{table*}[t]
\center
 \begin{tabular}{|c|c|c|c|c|}
\hline
 &   {\bf BP} & {\bf mod-CS} & {\bf weighted $\ell_1$}  & {\bf Reg-mod-BP}  \\
\hline
$p_{\text{exact}}(0.15m)$ & 0 &0.26 &0.26 & 0.57 \\
\hline
N-RMSE($0.15m$) & 0.967 & 0.152 & 0.152 &0.082 \\ \hline
$n_{\text{exact}}(1)$ & 0.4$m$ & 0.21$m$  & 0.21$m$ & 0.20$m$ \\
\hline
\end{tabular}
\vspace{-0.05in}
\caption{\small{Quantized signals and signal estimates: case 2. Recall that $k=|T|=26$. The expected sizes of $T_a$, $T_g$ and $T_b$ are $\E[|T_a|]=9.02$, $\E[|T_g|]=4.58$ and $\E[|T_b|]=21.42$.}}
\vspace{-0.1in}
\label{quantized_2}
\end{table*}

\begin{table*}[t]
\center
\begin{tabular}{|c|c|c|c|c|}
  \hline
     \ & {\bf BP} & {\bf mod-CS} & {\bf weighted $\ell_1$}  & {\bf Reg-mod-BP} \\
     \hline
     $p_{\text{exact}}(0.18$m$)$ & $0$ & $0.87$ & $0.87$ & $0.87$ \\
     \hline
     N-RMSE(0.18$m$) & 0.961 &  0.0175  & 0.0177  &  0.0123 \\
     \hline
    N-RMSE(0.11$m$)  & $1.05$ &  $0.179$  & $0.175$    & $0.0635$ \\
  \hline
  $n_{\text{exact}}(1)$ & $0.39m$ &  $0.21m$  & $0.21m$  & $0.21m$ \\
     \hline
\end{tabular}
\vspace{-0.05in}
\caption{\small{The non quantized case.}}
\vspace{-0.1in}
\label{not_quantized}
\end{table*}

For the quantized case,  $x$ was an $m=256$ length sparse vector with support size $|N|=0.1m=26$ and support estimate error sizes $u=|\Delta|=|\Delta_e|=0.1|N|=3$. We generated the matrix $A$ once as an $n \times m$ random Gaussian matrix (generate an $n \times m$ matrix with i.i.d zero mean Gaussian entries and normalize each column to unit $\ell_2$ norm). The following steps were repeated  $100$ times.
\ben
\item The support set, $N$, of size $|N|$, was generated uniformly at random from $[1,m]$. The support misses set, $\Delta$, of size $\sd$, was generated uniformly at random from the elements of $N$. The support extras set, $\Delta_e$, also of size $\sd$, was generated uniformly at random from the elements of $N^c$. The support estimate, $T = N \cup \Delta_e \setminus \Delta$ and thus $|T|=|N|=26$.
\item  We generated $x_{i} \stackrel{iid}{\sim} \text{discrete-uniform}(\pm 1)$ for $i \in N \cap T$; $x_{i} \stackrel{iid}{\sim} \text{discrete-uniform}(\pm 0.1)$ for $i \in \Delta$, and $x_{i}=0$ for $i \in N^c$. $x_{N \cap T}$ and $x_\Delta$ are also independent of each other.
    We generated $\muhat_{T} = x_{T} + \nu$ where  $\nu_i \stackrel{iid}{\sim} \text{discrete-uniform}(0,\pm \frac{\rho}{K},  \pm 2\frac{\rho}{K}, \dots  \pm \rho)$ for $i \in T \cap N$ and $\nu_i \stackrel{iid}{\sim} \text{discrete-uniform}(\pm \frac{\rho}{K},  \pm 2\frac{\rho}{K}, \dots  \pm \rho)$ for $i \in \Delta_e$. We used $\rho = 0.1$ and tried two choices of $K$. Notice that, for a given $K$, the number of equally likely values that $x_i - \muhat_i$ for $i \in T$ can take are roughly $2K+1$ ($2K$ when $i \in \Delta_e$). The constraint is active when $x_i - \muhat_i$ is equal to $\pm \rho$. Thus, the expected size of the active set is roughly $\frac{2}{2K+1}|T|$.

\item  We generated $y= Ax$. We solved reg-mod-BP given in (\ref{rmc}) with $\rho=0.1$;  BP given in (\ref{cs}); mod-CS given in (\ref{mc}); and weighted $\ell_1$ given in (\ref{wl1}) with various choices of $\gamma$: $[0.1 \ 0.05 \ 0.01 \ 0.001]$. We used the CVX optimization package, \url{http://www.stanford.edu/boyd/cvx/}, which uses primal-dual interior point method for solving the minimization problem.
\een
We computed $p_{\text{exact}}(n)$ as the number of times $\hat{x}$ was equal to $x$ (``equal" was defined as $\|\hat{x}-x\|_2/\|x\|_2 < 10^{-5}$) divided by $100$. For weighted $\ell_1$, we computed $p_{\text{exact}}(n)$ for each choice of $\gamma$ and recorded the largest one. This corresponded to $\gamma = 0.1$.
%
%
We tabulate results in Table \ref{quantized_1}. In the first row, we record $p_{\text{exact}}(0.15m)$ for all the methods, when using $K=2$. We also record the Monte Carlo average of the sizes of the active set $|T_a| = |T_{\text{a+}} \cup T_{\text{a-}}|$; of the good set, $|T_g| = |T_{\text{a+g}} \cup T_{\text{a-g}}|$ and of the bad set $|T_b| = k - |T_g|$.
In the second row, we record the normalized root mean squared error (N-RMSE). In the third row, we record $n_{\text{exact}}(1)$. In the next three rows, we repeat the same things with $K=5$.

As can be seen, $|T_g|$ is about half the size of the active set, $|T_a|$. As $K$ is increased, $|T_a|$ and hence $|T_g|$ reduces ($|T_b|$ increases) and thus $p_{\text{exact}}(0.15m)$ decreases and $n_{\text{exact}}(1)$ increases. Also, for mod-CS and weighted $\ell_1$, $p_{\text{exact}}(0.15m)$ is significantly smaller than for reg-mod-BP, while $n_{\text{exact}}(1)$ is larger.


Next, we simulated a more realistic scenario  -- the case of 3-bit quantized images (both $x$ and $\muhat$ take integer values between 0 to 7). Here again $m=256$, $|N|=0.1m=26$, and $u=|\Delta|=|\Delta_e|=0.1|N|=3$. The sets $N$, $\Delta$, $\Delta_e$ and $T$ were generated as before.  We generated $x_i \stackrel{iid}{\sim} \text{discrete-uniform}(3,4,\dots 7)$ for $i \in N \cap T$; $x_i \sim \text{discrete-uniform}(1,2)$ for $i \in \Delta$;  and $x_i = 0$ for $i \in N^c$. Also, $\muhat_{T} = \text{clip}(x_{T} + \nu)$ where $\nu_{i} \sim \text{discrete-uniform}(-2,-1,0,1,2)$ for $i \in T \cap N$; and  $\nu_{i} \sim \text{discrete-uniform}(-2,-1,1,2)$ for $i \in \Delta_e$. Also $\text{clip}(z)$ clips any value more than 7 to 7 and any value less than zero to zero.  Clearly, in this case $\rho = 2$. We record our results in  Table \ref{quantized_2}. Similar conclusions as before can be drawn.

Finally, we simulated the non-quantized case. We used $m=256$, $|N|=0.1m=26$, and $u=|\Delta|=|\Delta_e|=0.1|N|=3$. We generated $x_{i} \stackrel{iid}{\sim} \text{discrete-uniform}(\pm 1)$ for $i \in N \cap T$; $x_{i} \stackrel{iid}{\sim} \text{discrete-uniform}(\pm 0.1)$ for $i \in \Delta$, and $x_{i}=0$ for $i \in N^c$. The signal estimate, $\muhat_T = x_{T} + \nu$ where $\nu_i \stackrel{iid}{\sim} \text{uniform}(-\rho, \rho)$ with $\rho = 0.1$. We tabulate our results in Table \ref{not_quantized}.
Since $\nu$ is a real vector (not quantized), the probability of any constraint being active is zero. Thus, as expected, $p_{\text{exact}}$ and $n_{\text{exact}}$ are the same for reg-mod-BP and mod-CS and weighted $\ell_1$, though significantly better than BP. However, the N-RMSE for reg-mod-BP is significantly lower than that for mod-CS and weighted $\ell_1$ also, particularly when $n=0.11m$.

%% file: SecondRound_Revision_SupplemMaterial.tex
\section*{Supplementary Material}
\subsection{Proof of Theorem \ref{thm1}}
We construct a $w$ that satisfies the conditions of Lemma \ref{wcond} by first applying Lemma \ref{wbnd_iter0} and then applying Lemma \ref{wbnd} iteratively as explained below. Finally we define $w$ using (\ref{wdef}) below. At iteration zero, we apply Lemma \ref{wbnd_iter0} with $\Sp \equiv \sd$.
Lemma \ref{wbnd_iter0} can be applied because $k_b \le k$ and $\delta_{\sd} + \delta_{\st} + \theta_{\st,\sd}^2  < 1$ (holds because condition \ref{cond1} of the theorem holds). Thus, there exists a $w_1$ and an exceptional set $T_{d,1}$, disjoint with $T \cup \Delta$, of size less than $\Sp=\sd$, s.t.
\bea
{A_j}' w_1 & > & 0, \ \forall \ j \in T_{\text{a+g}}  \nn \\
{A_j}' w_1 & < & 0, \ \forall \ j \in T_{\text{a-g}}  \nn \\
{A_j}' w_1 \se 0,  \ \forall  \ j \in T_b   \nn \\
{A_j}' w_1 \se \text{sgn}(x_j), \ \forall \ j \in \Delta \nn \\
|T_{d,1}| &<&  \sd \nn \\
\|{A_{T_{d,1}}}'w_1\|_2 \sle a_{\stb}(\sd,\sd) \sqrt{\sd}  \nn \\
|{A_j}'w_1| \sle  a_{\stb}(\sd,\sd),  \ \forall j \notin T \cup \Delta \cup T_{d,1} \nn \\
\|w_1\|_2 \sle  K_{\stb}(\sd) \sqrt{\sd}
\label{iter_0}
\eea
At iteration $\iter > 0$, apply Lemma \ref{wbnd} with $T_{d} \equiv \Delta \cup T_{d,\iter}$ (so that $s \equiv 2\sd$), $c_j \equiv 0 \ \forall \ j \in \Delta$, $c_j \equiv {A_j}' w_\iter \ \forall \ j \in T_{d,\iter}$ and $\Sp \equiv \sd$. Call the exceptional set $T_{d,\iter+1}$.
Lemma \ref{wbnd} can be applied because $\delta_{2\sd} + \delta_{\st} + \theta_{\st,2\sd}^2 < 1$  (condition \ref{cond1} of the theorem).
 From Lemma \ref{wbnd}, there exists a $w_{\iter+1}$ and an exceptional set $T_{d,\iter+1}$, disjoint with $T \cup  \Delta \cup T_{d,\iter}$, of size less than $\Sp=\sd$, s.t.
\bea
{A_j}' w_{\iter+1} \se 0  \ \forall \ j \in T \nn \\
{A_j}' w_{\iter+1} \se 0, \ \forall \ j \in \Delta \nn \\
{A_j}' w_{\iter+1} \se {A_j}' w_\iter, \ \forall \ j \in T_{d,\iter} \nn \\
|T_{d,\iter+1}| &<& \sd \nn \\
\|{A_{T_{d,\iter+1}}}'w_{\iter+1}\|_2 \sle a_{\st}(2\sd,\sd)\|{A_{T_{d,\iter}}}'w_\iter\|_2  \nn \\
|{A_j}'w_{\iter+1}| \sle \frac{a_{\st}(2\sd,\sd)}{\sqrt{\sd}} \|{A_{T_{d,\iter}}}'w_\iter\|_2  \nn \\ && \forall j \notin T \cup \Delta \cup T_{d,\iter} \cup T_{d,\iter+1} \nn \\
\|w_{\iter+1}\|_2 \sle  K_{\st}(2\sd) \|{A_{T_{d,\iter}}}'w_\iter\|_2
\label{iter_n}
\eea
Notice that $|T_{d,1}| < \sd$ (at iteration zero) and $|T_{d,\iter+1}| < \sd$ (at iteration $\iter$) ensures that $|\Delta \cup T_{d,\iter}| < s=2\sd$ for all $\iter \ge 1$.%

The last three equations of (\ref{iter_n}), combined with the sixth equation of (\ref{iter_0}), simplify to
\bea
\|{A_{T_{d,\iter+1}}}'w_{\iter+1}\|_2 \sle a_{\st}(2\sd,\sd)^\iter a_{\stb}(\sd,\sd) \sqrt{\sd} \nn \\
\label{iter_n_2}
|{A_j}'w_{\iter+1}| \sle a_{\st}(2\sd,\sd)^\iter a_{\stb}(\sd,\sd),  \nn \\ && \forall j \notin T \cup \Delta \cup T_{d,\iter} \cup T_{d,\iter+1} \\
\|w_{\iter+1}\|_2 \sle  K_{\st}(2\sd) a_{\st}(2\sd,\sd)^{\iter-1} a_{\stb}(\sd,\sd) \sqrt{\sd} \nn \\
\label{wnbnd}
\eea
We can define
\bea
w \triangleq \sum_{\iter=1}^\infty (-1)^{\iter-1} w_\iter
\label{wdef}
\eea
Since $a_{\st}(2\sd,\sd) < 1$, $\|w_\iter\|_2$ approaches zero with $\iter$, and so the above summation is absolutely convergent, i.e. $w$ is well-defined. 

From the first four equations of (\ref{iter_0}) and first two equations of (\ref{iter_n}),
\bea
{A_j}' w & > & 0, \ \forall \ j \in T_{\text{a+g}}  \nn \\
{A_j}' w & < & 0, \ \forall \ j \in T_{\text{a-g}}  \nn \\
{A_j}' w \se  0,  \ \forall  \ j \in T_b \nn \\
{A_j}' w \se {A_j}' w_1 =  \text{sgn}(x_j), \ \forall \ j \in \Delta
\label{wcond12}
\eea
Consider ${A_j}' w = {A_j}' \sum_{\iter=1}^\infty (-1)^{\iter-1} w_\iter$ for some $j \notin T \cup \Delta$. If for a given $\iter$, $j \in T_{d,\iter}$, then ${A_j}' w_{\iter} = {A_j}' w_{\iter+1}$ (gets canceled by the $\iter+1^{th}$ term). If $j \in T_{d,\iter-1}$, then ${A_j}' w_{\iter} = {A_j}' w_{\iter-1}$ (gets canceled by the $\iter-1^{th}$ term). Since $T_{d,\iter}$ and $T_{d,\iter-1}$ are disjoint, $j$ cannot belong to both of them.
Thus,
\bea
{A_j}' w =  \sum_{\iter: j \notin T_{d,\iter} \cup T_{d,\iter-1} } (-1)^{\iter-1} {A_j}'w_\iter , \ \forall  j \notin T \cup \Delta
\eea
Consider a given $\iter$ in the above summation. Since $j \notin T_{d,\iter} \cup T_{d,\iter-1} \cup T \cup \Delta$, we can use (\ref{iter_n_2}) to get $|{A_j}'w_\iter| \le  a_{\st}(2\sd,\sd)^{\iter-1} a_{\stb}(\sd,\sd)$. Thus, for all $j \notin T \cup \Delta$,
\bea
| {A_j}' w | \sle \sum_{\iter: j \notin T_{d,\iter} \cup T_{d,\iter-1} } a_{\st}(2\sd,\sd)^{\iter-1} a_{\stb}(\sd,\sd) \nn \\
\sle \frac{a_{\stb}(\sd,\sd)}{1 - a_{\st}(2\sd,\sd)} 
\eea
Since $a_{\st}(2\sd,\sd) + a_{\stb}(\sd,\sd) < 1$ (condition \ref{cond2} of the theorem),
\bea
| {A_j}' w | < 1, \ \forall  j \notin T \cup \Delta
\label{wcond3}
\eea
Thus, from (\ref{wcond12}) and (\ref{wcond3}),  we have found a $w$ that satisfies the conditions of Lemma \ref{wcond}. From condition \ref{cond1} of the theorem,
$\delta_{\st+\sd} < 1$. Applying Lemma \ref{wcond}, the claim follows. $\blacksquare$

\subsection{Proof of Lemma \ref{wbnd}} 

Let $M = M(T)$.

The lemma assumes that  $\delta_{s} + \delta_{\st} + \theta_{\st,s}^2 < 1$. This means that (a) $\delta_{\st} < 1$ and so ${A_{T}}'A_{T}$ is positive definite; (b) $\delta_s < 1$ and so for any set $T_d$ of size $|T_d| \le s$, ${A_{T_d}}'{A_{T_d}}$ is positive definite; and (c) as we show next, for any set $T_d$ of size $|T_d| \le s$, ${A_{T_d}}'M A_{T_d} $ is also positive definite.
Notice that ${A_{T_d}}'M A_{T_d} = {A_{T_d}}'A_{T_d}  - {A_{T_d}}' A_{T}({A_{T}}'A_{T})^{-1}{A_{T}}'A_{T_d}$ which is the difference of two symmetric non-negative definite matrices. Let $B_1$ denote the first matrix and $B_2$ the second one. Use the fact that $\lambda_{\min}(B_1 - B_2) \ge \lambda_{\min}(B_1) + \lambda_{\min}(-B_2) =  \lambda_{\min}(B_1) - \lambda_{\max}(B_2)$ where $\lambda_{\min}(.), \lambda_{\max}(.)$ denote the minimum, maximum eigenvalue. Since $\lambda_{\min}(B_1) \ge (1-\delta_s)$ and $\lambda_{\max}(B_2) = \|B_2\| \le \frac{\|({A_{T_d}}' A_{T})\|^2}{1 - \delta_{\st}} \le \frac{\theta_{s,\st}^2}{1 - \delta_{\st}}$, thus
\bea
\lambda_{\min}({A_{T_d}}'M A_{T_d}) \ge {1-\delta_s - \frac{\theta_{s,\st}^2}{1 - \delta_{\st}} } > 0
\label{eq0_modcs}
\eea
(the last inequality holds because $\delta_{s} + \delta_{\st} + \theta_{\st,s}^2 < 1$). Thus, ${A_{T_d}}'M A_{T_d}$ is positive definite.

Since $M$ is a projection matrix, $MM' = M$, and so ${A_{T_d}}'M A_{T_d}  = {A_{T_d}}'M M' A_{T_d} $. Thus, from above, ${A_{T_d}}'M M' A_{T_d}$ is also positive definite. Thus, ${A_{T_d}}'M$ is full rank.

Any $\tw$ that satisfies ${A_T}'\tw = 0$ will be of the form
\bea
\tw = [I - A_{T}({A_{T}}'A_{T})^{-1}{A_{T}}'] \gamma : = M \gamma
\eea
We need to find a $\gamma$ s.t. ${A_{T_d}}'\tw = c$, i.e. ${A_{T_d}}'M \gamma = c$. Since ${A_{T_d}}'M$ is full rank, this system of equations has a solution (in fact, it has infinitely many solutions).
Let $\gamma = M'A_{T_d} \eta$. Then
$\eta = ({A_{T_d}}'M M'A_{T_d} )^{-1} c = ({A_{T_d}}'M A_{T_d} )^{-1} c$. This follows because $M M' = M^2 = M$ since $M$ is a projection matrix.
Thus,
\bea
\tw = MM'A_{T_d} ({A_{T_d}}'M A_{T_d} )^{-1} c = M A_{T_d} ({A_{T_d}}'M A_{T_d} )^{-1} c \ \ \ \
\label{def_tw}
\eea
Consider any set $\Tdp$ with $|\Tdp| \le \Sp$ disjoint with $T \cup T_d$. Then
\bea
\|{A_{\Tdp}}'\tw \|_2 
                  \sle \|{A_{\Tdp}}' M A_{T_d}\| \ \|({A_{T_d}}'M A_{T_d} )^{-1}\| \ \|c\|_2  \ \ \
\label{eq1_modcs}
\eea
Consider the first term from the right hand side (RHS) of (\ref{eq1_modcs}).
\bea
\|{A_{\Tdp}}' M A_{T_d}\| \sle \|{A_{\Tdp}}'A_{T_d}\| + \|{A_{\Tdp}}' A_{T}({A_{T}}'A_{T})^{-1}{A_{T}}' A_{T_d}\| \nn \\
                          \sle \theta_{\Sp,s} + \frac{\theta_{\Sp,\st} \ \theta_{s,\st}}{1 - \delta_{\st}}
\label{eq2_modcs}
\eea
This follows in a fashion exactly analogous to the derivation of the upper bound on the first term of (\ref{def_au}) in the proof of Lemma 2. 
Consider the second term from the RHS of (\ref{eq1_modcs}). Since ${A_{T_d}}'M A_{T_d}$ is positive definite,
\bea
\|({A_{T_d}}'M A_{T_d} )^{-1}\| \se \frac{1}{\lambda_{\min}({A_{T_d}}'M A_{T_d} )}
\eea
Using (\ref{eq0_modcs}),
\bea
\|({A_{T_d}}'M A_{T_d} )^{-1}\| \sle \frac{1}{1-\delta_s - \frac{\theta_{s,\st}^2}{1 - \delta_{\st}} }
\label{eq3_modcs}
\eea
Recall that the denominator is positive because we have assumed that $\delta_{s} + \delta_{\st} + \theta_{\st,s}^2 < 1$. Using (\ref{eq2_modcs}) and (\ref{eq3_modcs}) to bound (\ref{eq1_modcs}), we get that for any set $\Tdp$ with $|\Tdp| \le \Sp$,
\bea
\|{A_{\Tdp}}'\tw \|_2 \sle \frac{\theta_{\Sp,s}    + \frac{\theta_{\Sp,\st} \ \theta_{s,\st}}{1 - \delta_{\st}}}{ 1-\delta_s - \frac{\theta_{s,\st}^2}{1 - \delta_{\st}} } \|c\|_2  = a_{\st}(s,\Sp) \|c\|_2 \ \ \ \ \ \ \
 \label{def_a_2}
\eea
Notice that $a_{\st}(s,\Sp)$ is non-decreasing in $\st$, $s$, $\Sp$. 
Define an exceptional set, $E$, as%
\bea
E :=\{ j \in (T \cup T_d)^c : |{A_j}'\tw| > \frac{a_{\st}(s,\Sp) }{\sqrt{\Sp}} \|c\|_2 \}
\eea
Notice that $|E|$ must obey $|E| < \Sp$ since otherwise we can contradict (\ref{def_a_2}) by taking $\Tdp \subseteq E$.

Since $|E| < \Sp$ and $E$ is disjoint with $T \cup T_d$, (\ref{def_a_2}) holds for $\Tdp \equiv E$, i.e. $\|{A_{E}}'\tw \|_2 \le a_{\st}(s,\Sp) \|c\|_2$. Also, by definition of $E$, $|{A_j}'\tw| \le  \frac{a_{\st}(s,\Sp) }{\sqrt{\Sp}} \|c\|_2$, for all $j \notin T \cup T_d \cup E$.
Finally,
\bea
\|\tw\|_2 \sle \| M A_{T_d} ({A_{T_d}}'M A_{T_d} )^{-1} \| \ \|c\|_2 \nn \\
\sle \| M\| \ \|A_{T_d} \| \ \| ({A_{T_d}}'M A_{T_d} )^{-1} \| \  \|c\|_2   \nn \\
\sle  \frac{ \sqrt{1 + \delta_s} }{1-\delta_s - \frac{\theta_{s,\st}^2}{1 - \delta_{\st}} } \|c\|_2 = K_{\st}(s) \|c\|_2
\label{def_K_2}
\eea
since $\|M\| =  1$ (holds because $M$ is a projection matrix). Thus we have found a $\tw$ and a set $E$ that satisfy all conditions of the lemma.